\documentclass[12pt]{article}
\usepackage{amssymb}
\usepackage{epsfig,caption,graphicx,eepic,epic}
\usepackage{pstricks}
\usepackage{amsmath}
\usepackage{multido}

\begin{document}
\def\be{\begin{equation}}
\def\ee{\end{equation}}
\def\ba{\begin{eqnarray}} 
\def\ea{\end{eqnarray}}
\def\baa{\begin{eqnarray*}} 
\def\eaa{\end{eqnarray*}}
\def\nn{\nonumber}

\newcommand{\bbf}{\mathbf}
\newcommand{\rrm}{\mathrm}

\title {Study of the spectral properties of spin ladders in different representations via a renormalization procedure\\}

\author{Tarek Khalil$^{a,b}$
\footnote{E-mail address: tarek.khalil@liu.edu.lb}\\ 
and\\
Jean Richert$^{c}$
\footnote{E-mail address: richert@fresnel.u-strasbg.fr}\\ 
$^{a}$ Department of Physics, School of Arts and Sciences,\\
Lebanese International University, Beirut, Lebanon\\
$^{b}$ Department of Physics, Faculty of Sciences(V),\\
Lebanese University, Nabatieh, Lebanon\\
$^{c}$ Institut de Physique, Universit\'e de Strasbourg,\\
3, rue de l'Universit\'e, 67084 Strasbourg Cedex,\\ 
France} 

\date{\today}
\maketitle 
\begin{abstract}
We implement an algorithm which is aimed to reduce the dimensions of the Hilbert space
of quantum many-body systems by means of a renormalization procedure. We test the role and importance of different representations on the reduction process by working out and analyzing the spectral properties of 
strongly interacting frustrated quantum spin systems.
\end{abstract} 
\maketitle

PACS numbers: 02.70.-c; 03.65.-w; 05.10.Cc; 71.27.+a 

Keywords: Effective theories, renormalization, strongly interacting systems, quantum spin systems.

\section{Introduction.}

Most microscopic many-body quantum systems are subject to strong interactions which act 
between their constituents. In general, there exist no analytical methods to treat exactly 
strongly interacting systems apart from the assumption of trial wave functions able to 
diagonalize the Hamiltonian of integrable models, like the Bethe Ansatz $(BA)$ for specific 
one dimensional systems $(1D)$~\cite{bos}, the $BCS$ hypothesis which explains the supraconductivity and others like the N\'eel state, the Resonant Valence Bond ($RVB$) spin liquid 
states proposed by Anderson~\cite{ander1}, the Valence Bond Crystal states ($VBC$)\ldots . 
They are aimed to describe $2D$ systems but there remains the problem of their degree of realism, 
i.e. their ability to include the essentials of the interaction in strongly interacting systems
~\cite{lhuil}. The complexity of the structure of such systems leads to the diagonalization of 
the Hamiltonian numerically which must in general be performed in very large Hilbert space although 
the information of interest is restricted to the knowledge of a few low energy states generally 
characterized by collective properties. Consequently it is necessary to manipulate very large 
matrices in order to extract a reduced quantity of informations.\\

Non-perturbative techniques are needed. During the last decades a considerable amount of procedures 
relying on the renormalization group concept introduced by Wilson ~\cite{wil} have been proposed and 
tested. Some of them are specifically devised for quantum spin systems, like the Real Space 
Renormalization Group (RSRG)~\cite{mal,whi} and the Density Matrix Renormalization Group (DMRG)
~\cite{whi2,henk}.\\  

We propose here a non-perturbative approach which tackles this question ~\cite{khri}. 
The procedure consists of an algorithm which implements a step by step reduction of the 
size of Hilbert space by means of a projection technique. It relies on the renormalization 
concept following in spirit former work based on this concept~\cite{gla,mue,bek}.
Since the reduction procedure does not act in ordinary or momentum space but in Hilbert space, 
it is universal in the sense that it works for any kind of many-body quantum system.\\ 

The properties of physical systems can be investigated in different representations. In the present work which deals with frustrated spin ladders the common $SU(2)$-representation is confronted with the $SO(4)$-representation ~\cite{bohm,kika,kik} in order 
to work out the energies of the low-lying levels of the spectrum of these systems. The efficiency 
of one or the other representation in terms of the number of relevant basis states characterizing the ground 
state wave function is tested in connection with the reduction process.\\

The outline of the paper is the following. In section $2$ we present the formal developments leading to 
the secular equation in the reduced Hilbert space. Section $3$ is devoted to the application of the 
algorithm to frustrated quantum spin ladders with two legs and one spin per site. We analyze the 
outcome of the applied algorithm on systems characterized by different coupling strengths by means of 
numerical examples, in bases of states developed in the $SU(2)$ and $SO(4)$-representations and compare 
the results obtained in both cases. Conclusions and further possible investigations and developments are presented in section $4$.\\

\section{The reduction algorithm.}

\subsection{Reduction procedure and renormalization of the coupling strengths.}

We consider a system described by a Hamiltonian depending on a unique coupling strength $g$ which 
can be written as a sum of two terms 
\be
H = H_0 + g H_1 
\label{eq4} \  
\ee

The Hilbert space  ${\cal H}^{(N)}$ of dimension $N$ is spanned by an orthonormalized arbitrary
set of basis states 
$\left\{|\Phi_i\rangle, \,i=1,\cdots, N\right\}$. In this basis an eigenvector $|\Psi_l^{(N)}\rangle$ takes the form

\be
|\Psi_l^{(N)}\rangle = \sum_{i=1}^{N}  a_{li}^{(N)}(g^{(N)})|\Phi_i\rangle
\label{eq5} \  
\ee    
where the amplitudes $\{a_{li}^{(N)}(g^{(N)})\}$ depend on the value $g^{(N)}$ of
$g$ in  ${\cal H}^{(N)}$.\\

Using the Feshbach formalism~\cite{fesh} the Hilbert space may be decomposed into subspaces by 
means of the projection operators $P$ and $Q$, 
\be
{\cal H}^{(N)} = P{\cal H}^{(N)} + Q{\cal H}^{(N)}
\ee

In practice the subspace $ P{\cal H}^{(N)}$ is chosen to be of dimension 
$\mathrm{dim}\,P{\cal H}^{(N)}= N-1$ by elimination of 
one  basis state. The projected eigenvector $P|\Psi_l^{(N)}\rangle$ obeys the Schro\"edinger
equation  
\be
H_{eff}(\lambda_l^{(N)})P |\Psi_l^{(N)}\rangle =  \lambda_l^{(N)}
P |\Psi_l^{(N)}\rangle
\label{eq6} \ . 
\ee
where $H_{eff}(\lambda_l^{(N)})$ is the effective Hamiltonian which operates in the subspace  
$P{\cal H}^{(N)}$. It depends on the eigenvalue $\lambda_l^{(N)}$ which is the eigenenergy 
corresponding to $|\Psi_l^{(N)}\rangle$ in the initial space ${\cal H}^{(N)}$. The coupling 
strengths $g^{(N)}$ which characterizes the Hamiltonian $H^{(N)}$ in ${\cal H}^{(N)}$ is now 
aimed to be changed into $g^{(N-1)}$ in such a way that the eigenvalue in the new space 
${\cal H}^{(N-1)}$ is the same as the one in the complete space
\be
\lambda_l^{(N-1)} = \lambda_l^{(N)} 
\label{eq7} \  
\ee
The determination of $g^{(N-1)}$ by means of the constraint expressed by Eq.~(\ref{eq7}) 
is the central point of the procedure. It is the result of a renormalization procedure induced by the reduction of the vector space of dimension $N$ to $N-1$ which preserves the physical eigenvalue $\lambda_l^{(N)}$.\\

In the sequel $P|\Psi_1^{(N)}\rangle$ is chosen to be projection of the ground state eigenvector $|\Psi_1^{(N)}\rangle$ $(l=1)$ and $\lambda_1^{(N)} = \lambda_1^{(N-1)} = \lambda_1$ the corresponding eigenenergy. 
In ref.~\cite{khri} it is shown how $g^{(N-1)}$ can be obtained as a solution of an algebraic equation of the second degree. One gets explicitly a discrete quadratic equation 
\be
{a^{(N-1)}{g^{(N-1)}}^2 + b^{(N-1)}g^{(N-1)} +  c^{(N-1)}} = 0  
\label{eq8} \ 
\ee 
where 
\be
a^{(N-1)} = G_{1N} - H_{NN} F_{1N} 
\label{eq9} \ 
\ee

\be 
b^{(N-1)} =  a_{11}^{(N)} H_{NN}(\lambda_1^{(N)} - \alpha_1) +  F_{1N}
(\lambda_1^{(N)} - \alpha_N) 
\label{eq11} \ 
\ee

\be
c^{(N-1)} =   -a_{11}^{(N)}(\lambda_1^{(N)} - \alpha_1)
(\lambda_1^{(N)} - \alpha_N)  
\label{eq12} \ 
\ee
with
\be\nonumber
F_{1N} =  \sum_{i=1}^{N-1} a_{1i}^{(N)} \langle \Phi_1|H_1| \Phi_i \rangle  
\ee

\be\nonumber
G_{1N} =  H_{1N} \sum_{i=1}^{N-1} a_{1i}^{(N)} \langle \Phi_N|H_1| \Phi_i \rangle 
\ee

\be\nonumber
H_{ij} = \langle \Phi_i|H_1|\Phi_j\rangle  
\ee
and 
\be\nonumber
\alpha_i=\langle \Phi_i|H_0|\Phi_i\rangle, \,i=1,\cdots, N
\ee
\\
The reduction procedure is then iterated in a step by step decrease of the dimensions of the vector space, $N \mapsto N-1 \mapsto N-2 \mapsto...$ leading at each step $k$ to a coupling strength $g^{(N-k)}$ which can be given as the solution of a flow equation in a continuum limit description of the Hilbert space. The procedure can be generalized to Hamiltonians depending on several coupling constants which experience a renormalization during the reduction procedure under further constraints~\cite{kh}.\\

\subsection{Outline of the reduction algorithm.}
We sketch here the different steps of the procedure.\\

$1-$ Consider a quantum system described by an Hamiltonian $H^{(N)}$ which acts in an $N$-dimensional Hilbert space.

\vskip 0.2cm

$2-$ Compute the matrix elements of the Hamiltonian matrix $H^{(N)}$ in a definite basis of states $\{|\Phi_i\rangle, i=1,\ldots,N \}$. The diagonal matrix elements $\{\epsilon_i = \langle \Phi_i|H^{(N)}| \Phi_i \rangle\}$ are arranged either in increasing order with respect to the $\{\epsilon_i\}$ or in decreasing order of the absolute values of the ground state wave function amplitudes $|a_{1i}^{(N)}(g^{(N)})|$~\cite{khri1}.\\

\vskip 0.2cm

$3-$ Use the Lanczos technique to determine $\lambda_1^{(N)}$ and $|\Psi_1^{(N)}(g^{(N)})\rangle$~\cite{lanc1,lanc2}.\\

\vskip 0.2cm
$4-$ Fix $g^{(N-1)}$ as described in section 2.1. Take the solution of the algebraic second order  equation closest to  $g^{(N)}$ (see Eq.(\ref{eq8})).\\

\vskip 0.2cm
$5-$ Construct $H^{(N-1)} = H_0 + g^{(N-1)} H_1$ by elimination of the matrix elements of 
$H^{(N)}$ involving the state $|\Phi_N\rangle$.\\

\vskip 0.2cm
$6-$ Repeat procedures $3$, $4$ and $5$ by fixing at each step $k$ $\{k=1,\ldots,N-N_{min}\}$, $\lambda_1^{(N-k)}=\lambda_1^{(N)} = \lambda_1$. The iterations are stopped at $N_{min}$ corresponding to the limit of space dimensions for which the spectrum gets unstable.

\subsection{Some remarks.}

\begin{itemize}

\item The procedure is aimed to generate the energies of the low-energy excited states of strongly 
interacting systems and possibly the calculation of further physical quantities.
\item The implementation of the reduction procedure asks for the knowledge of $\lambda_1$ and 
the corresponding eigenvector $|\Psi_1^{(N-k)}\rangle$ at each step $k$ of the reduction 
process. The eigenvalue $\lambda_1$ is chosen as the physical ground state energy of the 
system. Eigenvalue and eigenvector can be obtained by means of the Lanczos algorithm
~\cite{henk,lanc1,lanc2} which is particularly well adapted to very large vector space dimensions. 
The algorithm fixes $\lambda_1^{(N-k)} =  \lambda_1^{(N)}$ and determines $|\Psi_1^{(N-k)}\rangle$ 
at each step.\\ 
\item The process does not guarantee a rigorous stability of the eigenvalue $\lambda_1$.  
$|\Psi_1^{(N-k-1)}\rangle$ which is the eigenvector in the space ${\cal H}^{(N-k-1)}$ 
and the projected state $P|\Psi_1^{(N-k)}\rangle$ of $|\Psi_1^{(N-k)}\rangle$ into ${\cal H}^{(N-k-1)}$ may differ from each other. As a consequence it may not be possible to keep  $\lambda_1^{(N-k-1)}$ rigorously equal to $\lambda_1^{(N-k)} = \lambda_1$. In practice the degree 
of accuracy depends on the relative size of the eliminated amplitudes $\{a_{1(N-k)}^{(N-k)}(g^{(N-k)})\}$.
This point will be tested by means of numerical estimations and further discussed below.\\  
\item The Hamiltonians of the considered ladder systems are characterized by a fixed total magnetic 
magnetization $M_{tot}$. We work in subspaces which correspond to fixed $M_{tot}$. The total 
spin $S_{tot}$ is also a good quantum number which defines smaller subspaces for fixed 
$M_{tot}$~\cite{low}. We do not introduce them here because projection procedures on $S_{tot}$ 
are time consuming. Furthermore we want to test the algorithm in large enough spaces although not 
necessarily the largest possible ones in this preliminary tests considered here. 

\end{itemize}

\section{Application to frustrated two-leg quantum spin ladders.}

\subsection{The model}

\subsubsection{SU(2)-representation.}

Consider spin-$1/2$ ladders~\cite{weih,lin2} described by Hamiltonians of the following type and shown in $Fig. 1$

\ba\label{eq13}
H^{(s,s)} &=& J_t\sum_{i=1}^{L} s_{i_1}s_{i_2} + J_l\sum_{<ij>}s_{i_1}s_{j_1} +
J_l\sum_{<ij>}s_{i_2}s_{j_2} + J_{1c}\sum_{(ij)}s_{i_1}s_{j_2}  
\\ \nn
&  & + J_{2c}\sum_{(ij)}s_{i_2}s_{j_1} 
\ea
 
\begin{figure}
\epsfig{file=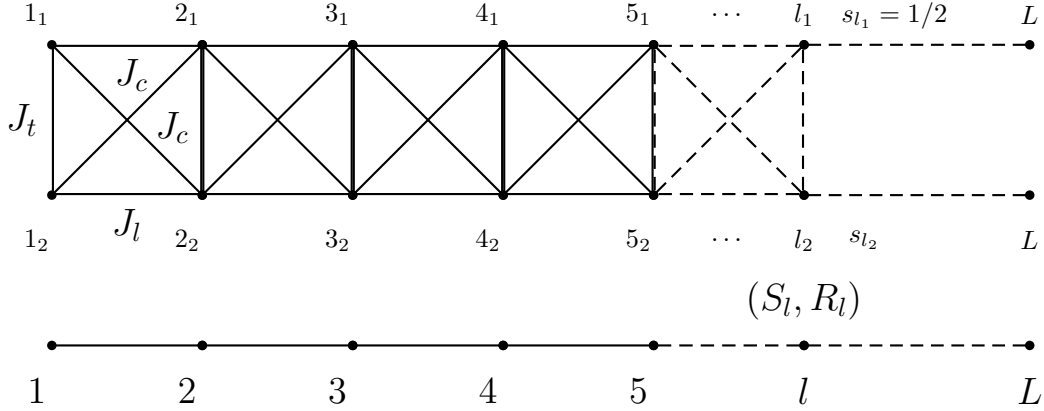}
\caption{Top: the original spin ladder. The coupling strengths are indicated as given in the text. 
Bottom: The ladder in the SO(4)-representation. See the text.}
\end{figure}

The indices $1$ or $2$ label the spin $1/2$ vector operators $s_{i_k}$ acting on the sites $i$ 
on both ends of a rung, in the second and third term $i$ and $j$ label nearest neighbours, here 
$j = i + 1$ along the legs of the ladder. The fourth and fifth term correspond to diagonal 
interactions between sites located on different legs, $j = i + 1$. $L$ is the number of sites on 
a leg (Fig. 1) where $J_{1c} = J_{2c} = J_c$. The coupling strengths $J_t, J_l, J_{c}$ are 
positive.\\

As stated above the renormalization is restricted to a unique coupling strength, see 
Eq.~(\ref{eq4}). It is implemented here by putting $H_0 = 0$ and $H^{(N)} = g^{(N)} H_1$ where 
$g^{(N)} = J_t$ and
\be
H_1 = \sum_{i=1}^{L} s_{i_1}s_{i_2} + \gamma_{tl} \sum_{<ij>}(s_{i_1}s_{j_1} +
s_{i_2}s_{j_2})  + \gamma_{c}\sum_{<ij>}(s_{i_1}s_{j_2} + s_{i_2}s_{j_1}) 
\label{eq14} \ . 
\ee
where $\gamma_{tl} = J_{l}/J_{t}$, $\gamma_{c} = J_{c}/J_{t}$. These quantities are kept constant 
and $g^{(N)} = J_t$  will be subject to renormalization during the reduction process.\\

One should point out that the renormalization does not change if one chooses another coupling parameter as a renormalizable parameter, here $J_l$ or $J_c$, because they are related to each other at the beginning of the reduction procedure by the ratios  $\gamma_{tl} = J_{l}/J_{t}$ and $\gamma_{c} = J_{c}/J_{t}$.\\
 
The basis of states $\{|\Phi_p\rangle\ ,  p=1,\ldots,N\}$ is chosen as

\be\nonumber
|\Phi_p\rangle = |1/2 ~~m_1,...,1/2 ~~m_i,...,1/2 ~~m_{2L}, \sum_{i=1}^{2L} m_i= M_{tot} = 0\rangle  
\ee
with $\{m_i = +1/2, -1/2\}$.

\subsubsection{SO(4)-representation.}

Different choices of bases may induce a more or less efficient reduction procedure depending on the strength of the coupling constants $J_{t}, J_{l}, J_{c}$. This point is investigated here by choosing also a basis of states which is written in an $SO(4)$-representation.\\
 
We replace $(s_{i1},s_{i2})$ corresponding to dimers by $({S_i},{R_i})$. By means of a spin rotation ~\cite{kika,kik}
\be
 s_{i_1} = \frac{1}{2} (S_i + R_i)
\label{eq15} \ .
\ee
\be
 s_{i_2} = \frac{1}{2} (S_i - R_i)
\label{eq16} \ .
\ee

The Hamiltonian, Eq.~(\ref{eq13}), can be expressed in the form\\
 
\be
H^{(S,R)} = \frac{J_t}{4}\sum_{i=1}^{L} (S_{i}^{2} -  R_{i}^{2}) + J_1\sum_{<ij>}S_i S_j + 
J_2\sum_{<ij>}R_i R_j
\label{eq17} \   
\ee
The structure of the corresponding system is shown in the lower part of Fig. 1.
Here  $J_{1}=(J_{l}+J_{c})/2$, $J_{2}=(J_{l}-J_{c})/2$  and as before $J_{1c}=J_{2c}=J_{c}$.
The components $S_{i}^{(+)}, S_{i}^{(-)}, S_{i}^{(z)}$ and $R_{{i}}^{(+)}, R_{{i}}^{(-)},
R_{{i}}^{(z)} $ of the vector operators $S_{i}$ and  $R_{i}$ are the $SO(4)$ group generators 
and $<ij>$ denotes nearest neighbour indices.\\

In this representation the states $\{|S_{i}M_{i}\rangle\}$ are defined as  
\ba\nonumber
|S_i M_i \rangle = \sum_{m_1,m_2} \langle 1/2 ~~m_1 ~~1/2 ~~m_2|S_i M_i  \rangle
|1/2 ~~m_1\rangle_{i}  |1/2 ~~m_2\rangle_{i}
\ea
along a rung are coupled to $S_i = 0$ or $S_i = 1$. Spectra are constructed in this representation 
as well as in the $SU(2)$-representation and the states $\{|\Phi_p\rangle\}$
take the form

\be\nonumber
|\Phi_p\rangle = |S_1 M_1,...,S_i M_i,...,S_{L} M_{L}, \sum_{i=1}^{L} M_i= M_{tot} = 0\rangle  
\ee

\subsection{Test observables}
In order to quantify the accuracy of the procedure we introduce different test quantities in order to estimate deviations between ground state and low excited state energies in 
Hilbert spaces of different dimensions. The stability of low-lying states can be estimated by means 
of    

\ba\label{18}
p(i) = |\frac{(e_i^{(N)}-e_i^{(n)})}{e_i^{(N)}}| \times 100 & with& i=1,\ldots,4
\ea  
where $e_i^{(n)} = \lambda_i^{(n)}/2L$ with $n=(N-k)$ corresponds to the energy per site at the $i$th physical state starting from the ground state at the $k$th iteration in Hilbert space. This quantity provides a percentage of loss of accuracy of the eigenenergies in the different reduced spaces.\\

A global characterization of the ground state wavefunction in different representations can also 
be given by the entropy per site in a space of dimension $n$

\ba\label{19}
s = - \frac{1}{2L} \sum_{i=1}^{n}{P_i}ln{P_i} & with & P_i = |\langle\Phi_i^{(n)}|\Psi_1^{(n)}
\rangle|^{2} = |a_{1i}^{(n)}|^{2}
\ea
which works as a global measure of the distribution of the amplitudes $\{a_{1i}^{(n)}\}$ in the 
physical ground state.\\

In the remaining part we work out the spectra of different systems and compare results obtained
in the two representations introduced above.\\

\subsection{Spectra in the SU(2)-representation.}

Results obtained with an $SU(2)$-representation basis of states are shown in $Figs. (2-3)$.\\ 

\subsubsection{First case: $L$= 6, $J_t$=15, $J_l$=5, $J_{c}$=3}
 
We choose the basis states in the framework of the $M$-scheme corresponding to subspaces with 
fixed values of the total projection of the spin of the $\{|\Phi_i\rangle\}$, $M_{tot}=0$.\\ 

In the present case $J_t > J_l, J_{c}$.   The dimension of the subspace is reduced step by step as explained above starting from $N=924$. As stated in section $2.2$ the basis states  $\{|\Phi_i\rangle\}$ are ordered with increasing energy of their diagonal matrix elements $\{\epsilon_i\}$ and eliminated starting from the state with largest energy $\epsilon_N$.\\ 

\begin{figure}[ht]
\includegraphics{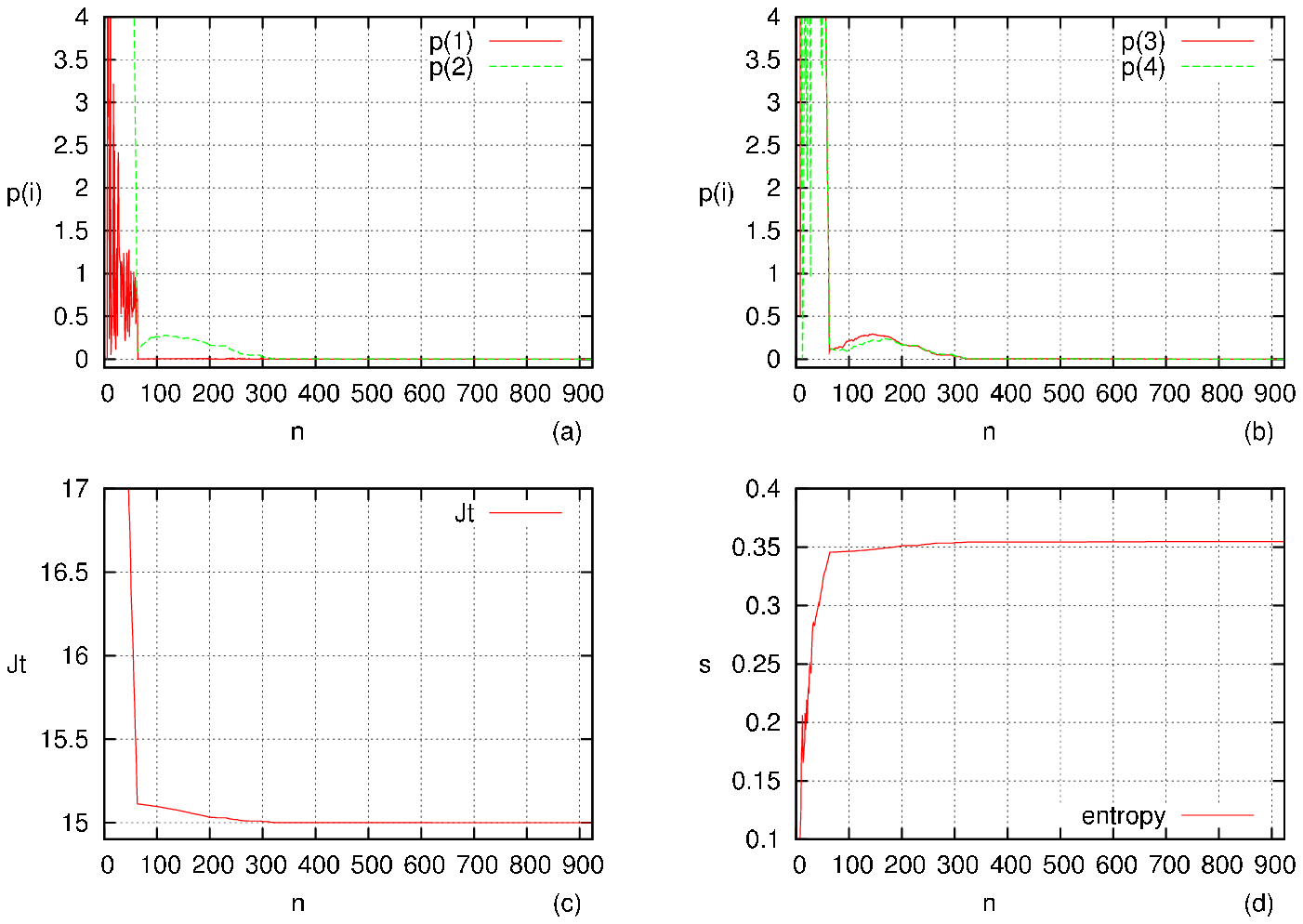}
\caption{$SU(2)-representation$. $n$ is Hilbert space dimension. $L=6$ sites along a leg. $J_t=15$, $J_l=5$, $J_{c}=3$}
\label{fig2}
\end{figure}

Deviations of the energies of the ground and first excited states from their initial values at $N=924$ can be seen in Figs.(\ref{fig2}a-b) where the $p(i)$'s defined above represent these deviations in terms of percentages.
As seen in Fig.(\ref{fig2}a) the ground state of the system stays stable down to $n\sim50$ where $n$ is the dimension of the reduced space. The coupling constant $J_t$ does not move either down to $n\sim 300$, see Fig.(\ref{fig2}c). Figs.(\ref{fig2}a-b) show the evolution of the first excited states which follows the same trend as the ground state.\\

For $n \leq 50$ the spectrum gets unstable, the renormalization of the coupling constant can 
no longer correct for the energy of the lowest state. Indeed the coupling constant $J_t$ increases
drastically as seen in Fig.(\ref{fig2}c). The reason for this behaviour can be found in 
the fact that at this stage the algorithm eliminates states which have an essential component in the 
state of lowest energy. The same message can be read on Fig.(\ref{fig2}d), the drop in the entropy  
per site $s$ is due to the elimination of sizable amplitudes $\{a_{1i}\}$.\\

\subsubsection{Second case: $L$= 6, $J_t$=5.5, $J_l$=5, $J_{c}$=3}
 
\begin{figure}[ht]
\includegraphics{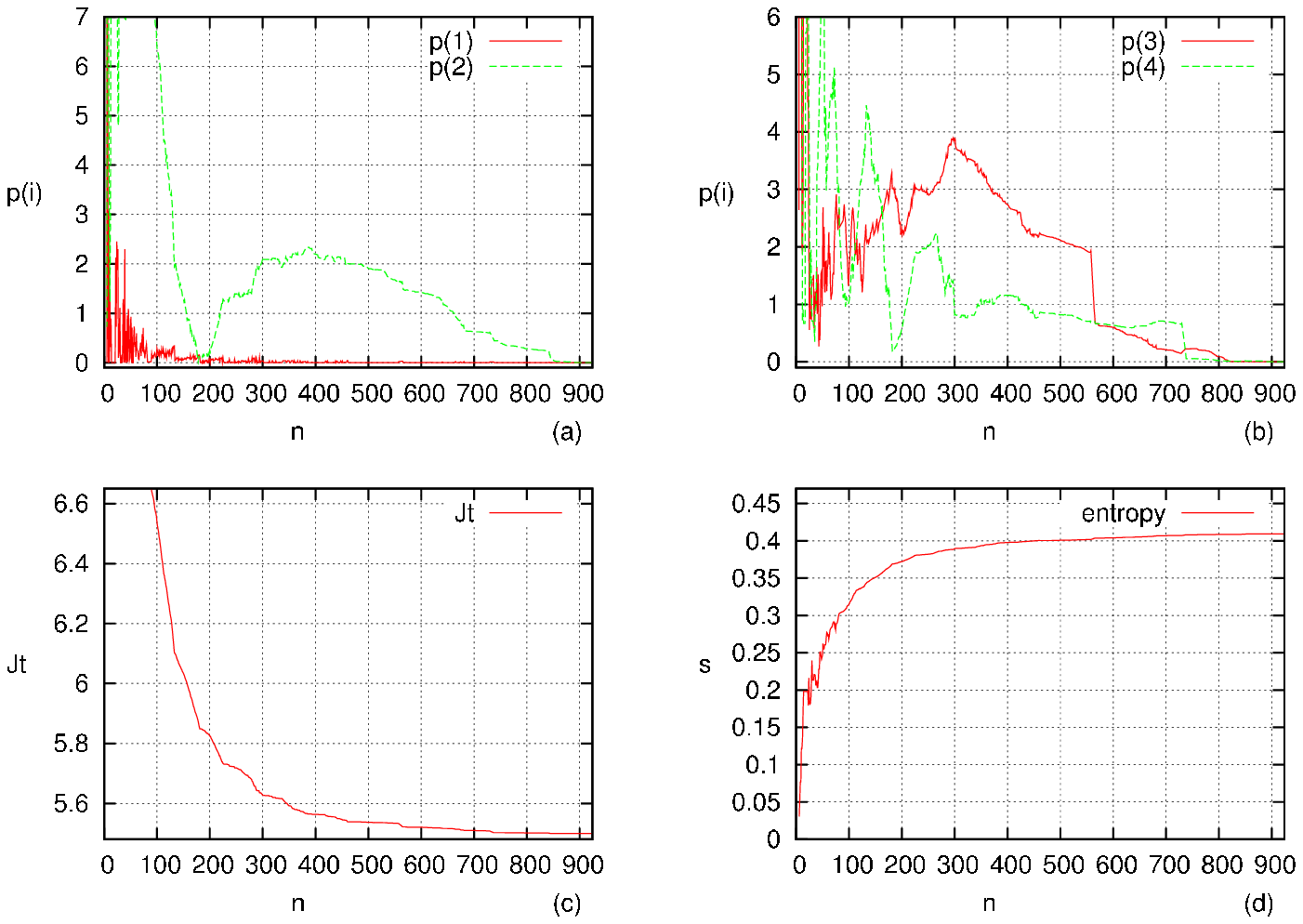}
\caption{$SU(2)-representation$. $n$ is Hilbert space dimension. $L=6$ sites along a leg.
$J_t=5.5$, $J_l=5$, $J_{c}=3$}
\label{fig3}
\end{figure}

Contrary to the former case the coupling constant $J_t$ along rungs is now of the order of strength as $J_l, J_{c}$. Results are shown in Fig.(\ref{fig3}). The lowest energy state is now stable down to $n \sim 100$. This is also reflected in the behaviour of the excited states which move appreciably for $n \leq 200$. Fig.(\ref{fig3}c) shows that the coupling constant $J_t$ starts to increase sharply between $n=300$ and $n=200$. It is able to stabilize the excited states down to about $n=200$ and the ground state down to $n=70$. The instability for $n\leq70$ reflects in the evolution of the $p(i)$'s, Figs.(\ref{fig3}a-b) which get of the order of a few percent. The entropy Fig.(\ref{fig3}d) follows the same trend.\\

Comparing the two cases above and particularly the entropies Fig.(\ref{fig3}d) and Fig.(\ref{fig2}d) one sees that the stronger $J_t$ the more the amplitude strength of the ground state wavefunction is concentrated in a smaller number of basis state components. The elimination of sizable components of the wavefunction leads to deviations which can be controlled down to a certain limit by means of the renormalization of $J_t$. One sees that large values of $J_t$ favour a low number of significative components in the low energy part of the spectrum in a $SU(2)$- representation.\\

\subsubsection{Remark}

In Figs.(\ref{fig2}a-b) it is seen that the ground and first excited states show "bunches" of energy fluctuations.  In the ground state the peaks are intermittent, they appear and disappear during the space dimension reduction process. They are small in the case where $J_t = 15$ but can grow with decreasing $J_t$ as it can be observed in Figs.(\ref{fig3}a-b) for $J_t = 5.5$. The subsequent stabilization of the ground state energy following such a bunch shows the effectiveness of the coupling constant renormalization which acts in a progressively reduced and hence incomplete basis of states.\\

These bunches of fluctuations are correlated with the change of the number of relevant amplitudes 
(i.e. amplitudes larger than some value $\epsilon$ as explained in the caption of Fig.(\ref{fig5})) 
during the reduction process.\\

Consider first the case where $J_t > J_l, J_c$. One notices in the caption of Fig.(\ref{fig5}a) that down to $n\sim300$ the number of relevant amplitudes defined in Fig.(\ref{fig5}) stays stable like the ratios $\{p(i)\}$ in Figs.(\ref{fig2}a-b). For $158<n<300$ these ratios change quickly. 
A bunch of fluctuations appears in this domain of values of $n$ as seen in Figs.(\ref{fig2}a-b) and correspondingly the number of relevant amplitudes decreases steeply. For $60 < n \leq158$ the ratios $\{p(i)\}$ stay again stable as well as the number of relevant amplitudes. The $\{p(i)\}$ in Figs.(\ref{fig2}a-b) almost decrease back to their initial values. The analysis shows that these 
bunches of fluctuations 
signal the local elimination of relevant contributions of basis states to the physical states in the 
spectrum. The stabilization of the spectra which follows during the elimination process shows that 
renormalization is able to cure these effects.\\

In the case where $J_t \sim J_l, J_{c}$ shown in Fig.(\ref{fig5}b) the relevant and irrelevant amplitudes move continuously during the reduction process and the corresponding $\{p(i)\}$ do no longer decrease to the values they showed before the appearance of a bunch of energy fluctuations as seen in Figs.(\ref{fig3}a-b). It signals the fact that the coupling renormalization is no longer able to compensate for the reduction of the Hilbert space dimensions.\\

One should mention that the evolution of the spectrum depends on the initial size of Hilbert space. The larger the initial space the larger the ratio between the initial number of states and the number of states corresponding to the limit of stability of the spectrum, see ref.~\cite{kh} for explicit numerical examples.

\subsection{Spectra in the SO(4)-representation}

The reduction algorithm is now applied to the system described by the Hamiltonian $H^{(S,R)}$ given by Eq.~(\ref{eq17}) with a basis of states written in the $SO(4)$-representation. Like above we consider two cases corresponding to large and close values of $J_t$ relative to the strengths of the other coupling parameters.

\subsubsection{Reduction test for $L=6$, $J_t$ = 15, 5.5 and $J_l$=5, $J_{c}$=3}
 
\begin{figure}[ht]
\includegraphics{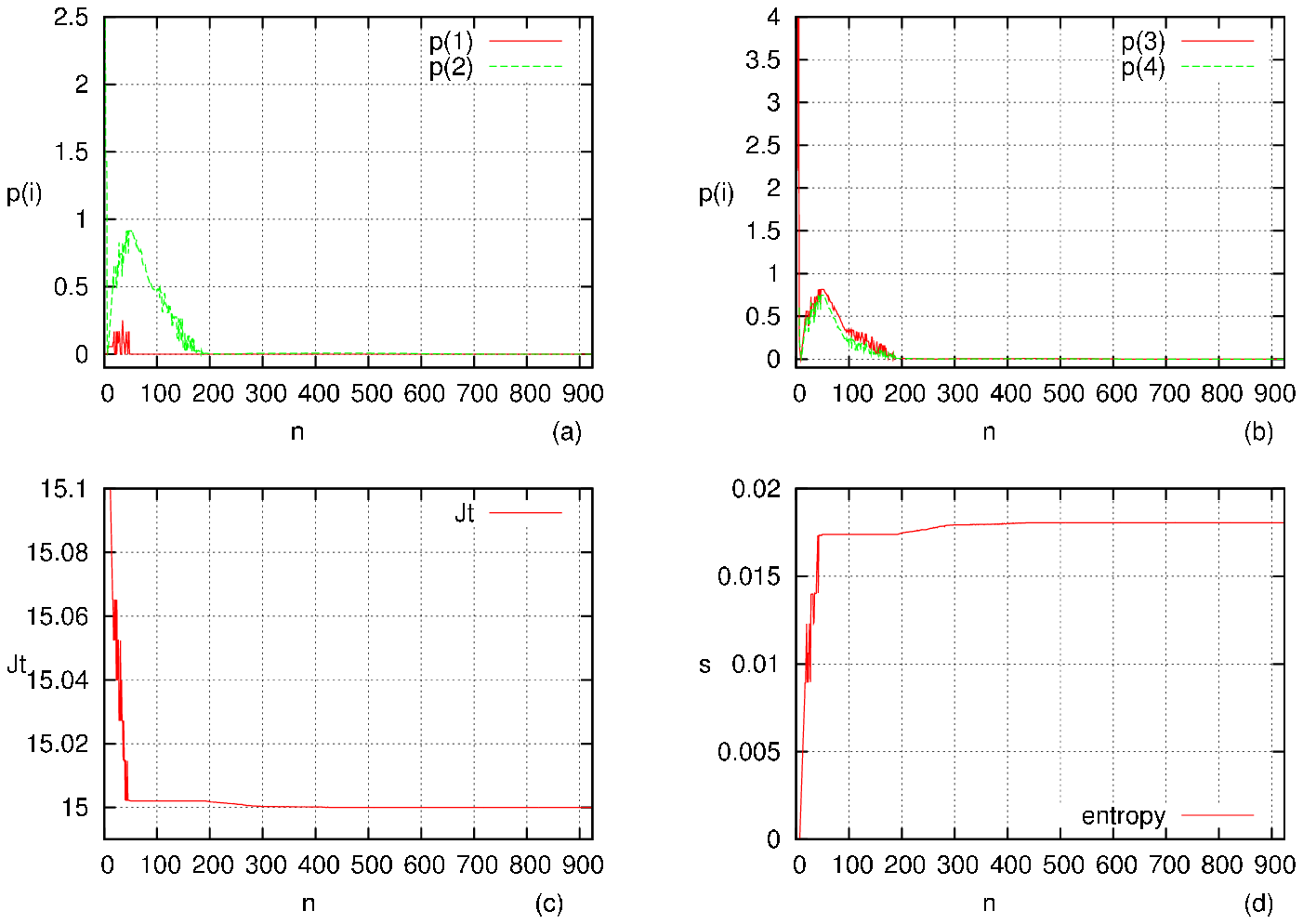}
\caption{$SO(4)-representation$. $n$ is Hilbert space dimension. $L=6$ sites along the chain. $J_t=15$, $J_l=5$, 
$J_{c}=3$}
\label{fig7}
\end{figure}

\begin{figure}[ht]
\includegraphics{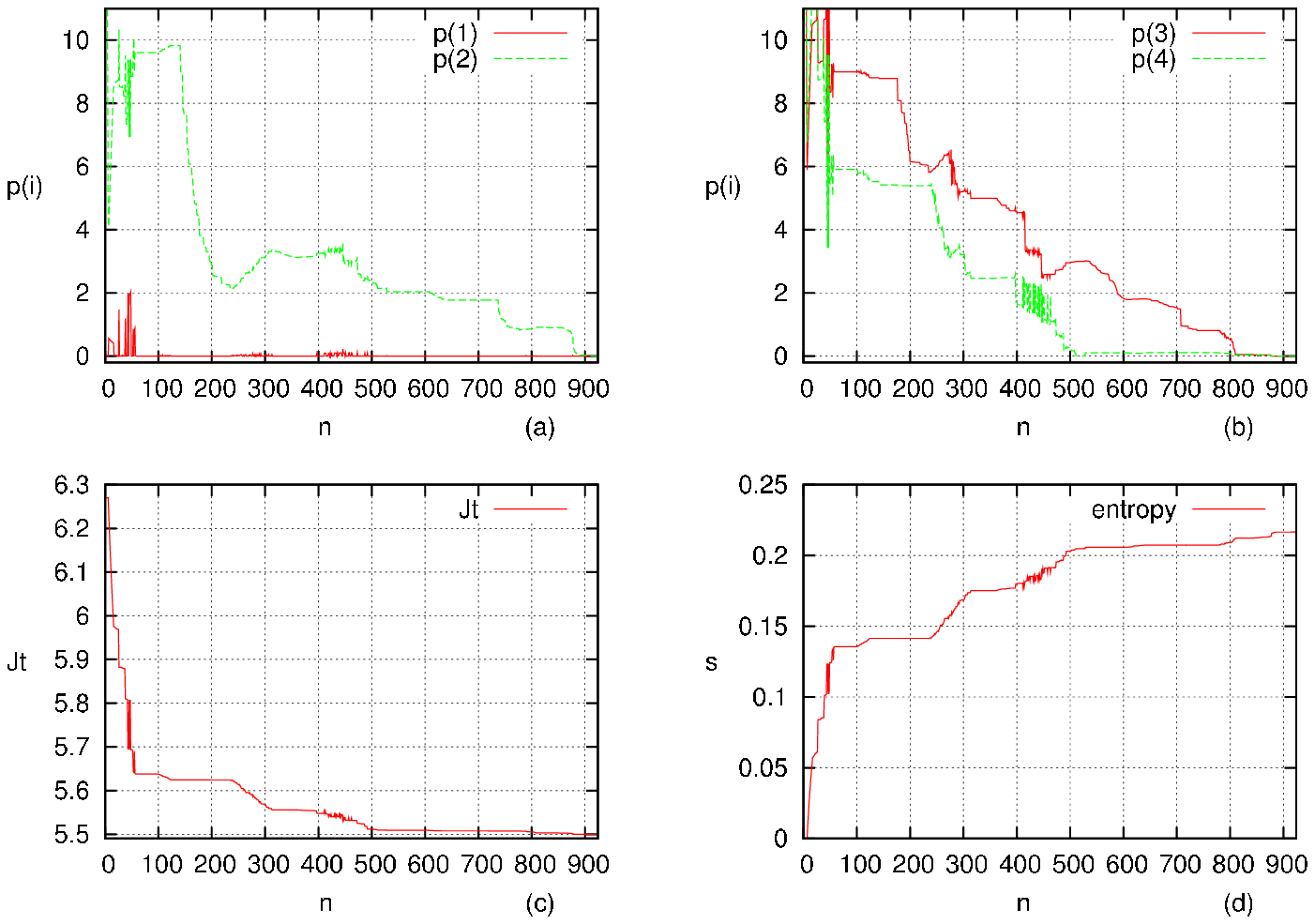}
\caption{$SO(4)-representation$. $n$ is Hilbert space dimension. $L=6$ sites along the chain. $J_t=5.5$, $J_l=5$, $J_{c}=3$}
\label{fig9}
\end{figure}

Figs.(\ref{fig7}-\ref{fig9}) show the behaviour of the spectrum for a system of size $L=6$. A large value of 
$J_t$, ($J_t = 15$), favours the dimer structure along rungs in the lowest energy state and 
stabilizes the spectrum down to small Hilbert space dimensions. This effect is clearly seen in 
Fig.(\ref{fig7}a), the ground state is very stable. The excited states are more affected, see  
Fig.(\ref{fig7}b), although they do not move significantly. The 
renormalization of the coupling strength $J_t$ starts to work for $n \simeq 50$, Fig.(\ref{fig7}c).\\ 
 
\begin{figure}[ht]
\includegraphics{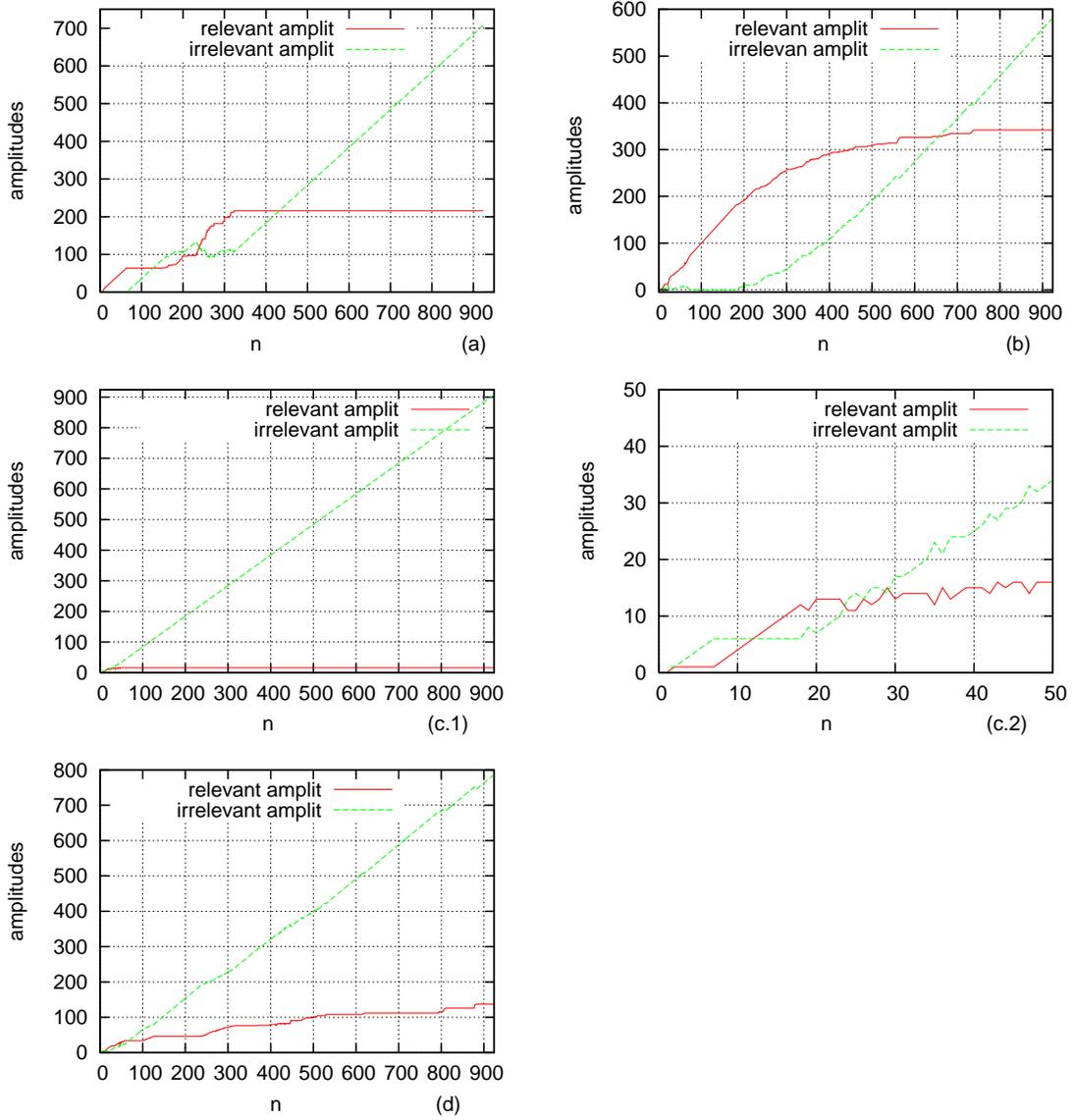}
\caption{$n$ is the dimension of the Hilbert space. $Amplitudes$ show the number of relevant -irrelevant amplitudes in the ground state eigenfunction (relevant-irrelevant amplit in the figure). Relevant amplitudes are those for which $\{|a_{1i}|>\epsilon$, (here $\epsilon = 10^{-2}$), $i=1,\ldots,n\}$. The number of sites along a leg is $L=6$. $SU(2)-representation$: (a) corresponds to $J_t=15$, $J_l=5$, $J_{c}=3$. (b) corresponds to $J_t=5.5$, $J_l=5$, $J_{c}=3$. $SO(4)-representation$: (c.1) and (c.2) correspond to $J_t=15$, $J_l=5$, $J_{c}=3$. (d) corresponds to $J_t=5.5$, $J_l=5$, $J_{c}=3$.} 
\label{fig5}
\end{figure}

The situation changes progressively with decreasing values of $J_t$. Figs.(\ref{fig9}) show the case where $J_t=5.5$. The ground state energy experiences sizable bunches of fluctuations like in the $SU(2)$-representation, but much stronger than in this last case. The same is true for the excited states which is reflected through all the quantities shown in Figs.(\ref{fig9}a-b), in particular $J_t$, Fig.(\ref{fig9}c). The entropy Figs.(\ref{fig7}d,\ref{fig9}d) follows the same trend like in the $SU(2)$-representation by changing the values of $J_{t}$ from $15$ to $5.5$. The analysis of the bunches of energy fluctuations through the number of relevant-irrelevant amplitudes is shown for $J_t=15$, $J_l=5$, $J_{c}=3$ in Fig. (\ref{fig5}c.1-c.2) and for $J_t=5.5$, $J_l=5$, $J_{c}=3$ in Fig. (\ref{fig5}d).\\ 

The results show that the renormalization procedure is quite sensitive to the representation chosen in Hilbert space. It is expected that essential components of the ground state wavefunction get eliminated early during the process when the rung coupling gets of the order of magnitude or smaller than the other coupling strengths.\\
 
By comparing the two representations in Fig(\ref{fig5}), one notices that the stability of the low-energy properties of the system in the reduced Hilbert space is well characterized by the number of relevant-irrelevant amplitudes as defined in the caption of Fig.(\ref{fig5}) and the distribution of the amplitudes in Hilbert space, see ref.~\cite{khri1}.\\ 

\subsection{Summary}

The present results lead to two correlated remarks. The efficiency of the algorithm is different
in different sectors of the coupling parameter space. In the case of the frustrated ladders 
considered here the algorithm is the more efficient the stronger the coupling between rung sites 
$J_t$. Second, this behaviour is strongly related to the representation in which the basis 
of states is defined. The $SU(2)$-representation leads to a structure of the wave functions (i.e. 
the size of the amplitudes of the basis states) which is very different from the one obtained in 
the $SO(4)$-representation. For large values of $J_t$ the spectrum is more stable in the $SO(4)$-representation. For small values of $J_t$ the stability is better realized in the $SU(2)$- representation. Finally, 
in the regime where $J_t > J_l, J_{c}$, one observes that the reduction procedure is the more 
efficient the closer $J_l$ to $J_{c}$. This effect can be understood and related to previous 
analytical work in the $SO(4)$-representation~\cite{jr2}.\\
 
\section{Conclusions and outlook.}

In the present work we tested and analysed the outcome of an algorithm which aims to reduce the 
dimensions of the Hilbert space of states describing strongly interacting systems. The reduction is 
compensated by the renormalization of the coupling strengths which enter the Hamiltonians of the 
systems. By construction the algorithm works in any space dimension and may be applied to the study 
of any microscopic $N$-body quantum system. The robustness of the algorithm has been tested on 
frustrated quantum spin ladders.\\

The analysis of the numerical results obtained in applications to quantum spin ladders leads to the 
following conclusions.  

\begin{itemize} 

\item The present numerical applications are essentially realistic tests of the renormalization 
algorithm restricted to rather small Hilbert spaces which do not necessarily necessitate the use of 
a Lanczos procedure. We introduced this procedure in order to be able to extend our present work to much 
larger systems for which ordinary full diagonalization cannot be performed. We tested it sucessfully 
by comparing the outcome with that of complete diagonalization. Preliminary extended calculations have 
also been performed on higher dimensional ladders. We expect to present and discuss the physics of their 
application to frustrated systems in further work.\\ 

\item The stability of the low-lying states of the spectrum in the course of the reduction procedure 
depends on the relative values of the  coupling strengths. The ladder favours a dimer structure 
along the rungs, i.e. stability is the better the larger the transverse coupling strength $J_t$.  
This is the reason why the $SO(4)$-representation is favoured when compared to the $SU(2)$- representation in this 
case. It leads to a more efficient basis of states as commented below (see next paragraph).\\

\item The efficiency of the reduction procedure depends on the representation frame in which the basis of 
states is defined. It appears clearly that the evolution of the spectrum described in an $SU(2)$- 
representation is significantly different from the evolution in an $SO(4)$-representation. This is understandable since 
different representations partition Hilbert space in different ways and favour one or the other representation
depending on the relative strengths of the coupling constants. It is always appropriate to work in 
a basis of states whose symmetry properties are closest to the symmetry properties of the physical system 
so that the contributions of the non-diagonal couplings are the smallest. The choice of an optimal basis 
may not always be evident. In the case of the ladder we consider here the $SO(4)$-representation.

\item Local spectral instabilities appearing in the course of the reduction procedure are correlated with 
the elimination of basis states with sizable amplitudes in the ground state wavefunction.
One or another representation can be more efficient on the reduction process for a given set of coupling parameters because it 
leads to physical states in which the weight on the basis states is concentrated in a different number 
of components. This point is strongly related to the correlation between quantum entanglement and 
symmetry properties which have been under intensive scrutiny, see f.i. ~\cite{kor} and refs. therein.  
  
\end{itemize} 

Further points are worthwhile to be investigated:

\begin{itemize} 

\item  Extension to larger ladders and systems of higher space dimensions with the help of more sophisticated numerical algorithms ~\cite{whith,caur}. 

\item Extension of the renormalization procedure to systems at finite temperature ~\cite{jr} and more than one coupling constant renormalization~\cite{kh}.

\end{itemize}

\newpage

\end{document}